# Gyroscopic Stability of Open Tipped Match Style Rifle Bullets

Elya R. Courtney and Michael W. Courtney

BTG Research, Baton Rouge, Louisiana

Michael_Courtney@alum.mit.edu

**Abstract:** Earlier work has produced formulas for predicting stability of rifle bullets of near uniform density and also for plastic-tipped rifle bullets. These formulas have been shown to be accurate within 5%. However, the original Miller stability formula for metal bullets of near uniform density underestimates the stability of match style open tipped rifle bullets having a significant empty volume in the tip. This paper presents a new formula for accurately estimating the stability of these open tipped match style rifle bullets from parameters easily obtained such as the bullet mass, length, and depth of the empty space in the tip. The formula is tested by observations of whether bullets tumble and by measuring the aerodynamic drag vs. predicted stability of several bullets over a range of stabilities. In addition to confirmation in several test cases with original firing data, the new stability formula is validated by comparison with BRL spark range stability data for the 168 grain Sierra International in .308 to an accuracy of 2% for supersonic velocities (> M1.2).

**Introduction**

Prior to publication of the Miller Twist Rule (Miller, 2005), the most common methods for estimating bullet stability were the relatively inaccurate Greenhill formula and the expensive computer modeling program called PRODAS. The modeling program is known to be accurate and is widely used by government laboratories and bullet companies. Bullet companies usually publish a minimum recommended twist rate, usually a twist that will ensure bullet stability under the most dense atmospheric conditions a hunter or recreational shooter is likely to encounter, with some margin of error so that no shooter is likely to ever report evidence of bullet tumbling as long as the minimum twist rate is used.

Professionals and recreational shooters often desire a way to estimate bullet stability when shooting under varying environmental conditions, using a rifle and bullet combination outside of the designers' original intent (plinking loads, subsonic applications, heavier bullets, etc.), trying to enhance terminal performance with early yaw in tissue, or trying to improve accuracy by reducing the probability of tumbling when shooting through brush. Don Miller's original twist rule provided shooters with an empirical twist rule that was more accurate than the Greenhill formula. Consequently, in spite of the assumption of constant bullet density, the twist rule was incorporated into several ballistics calculators and soon came into wide use. Later, a modified twist formula was published to more accurately predict the gyroscopic stability of plastic-tipped bullets. (Courtney and Miller, 2012a). Because of the assumption of constant bullet density, the original twist rule underestimated the stability of plastic tipped bullets and also underestimated the stability for open tipped match style bullets that have a significant volume of empty space in the tip.

Don Miller and Michael Courtney published an improved formula predicting the stability of plastic tipped bullets in early 2012 (Courtney and Miller, 2012a; Courtney and Miller, 2012b), and the authors of the present study later published an experimental paper further validating the original twist rule and improved formula for plastic-tipped bullets in early 2013 (Courtney and Courtney, 2013). This left the open tipped match style bullet as a common bullet design without an available formula to estimate the gyroscopic stability from easily obtainable parameters such as bullet mass, diameter, and length. In addition to their applications in shooting matches, open tipped match bullets have a number of military and law enforcement applications, due to their excellent accuracy (Roberts, 2008; Parks, 2012).

The present study presents an improved stability formula for estimating the gyroscopic stability of open tipped match style bullets and shows that the formula is accurate by observations of whether or not bullets tumble, experimentally studying ballistic coefficient vs. gyroscopic stability (Sg) for several bullets, and comparison with experimentally determined Sg

# Gyroscopic Stability of Open Tipped Match Style Rifle Bullets

values (from the BRL spark range) for the 168 grain Sierra International bullet in .308.

**Formula**

The development of the formula for gyroscopic stability of open tipped match (OTM) bullets was motivated by the notion that the solution to new problems often lies between the answers to other problems that are already solved. The formula for Sg of plastic tipped bullets (Courtney and Miller, 2012a; Courtney and Miller, 2012b) employed the idea that there were two important lengths in the original formula: 1) the total length of the bullet because it was related to the center of pressure, and the pressure of the air flow depended on the surface of the bullet and not on the material density; 2) the length of the higher density portion of the bullet because it was related to the center of mass. In the earlier works, these were referred to as the total length and the metal length, respectively. It was convenient to be able to express the stability formula in terms of parameters that could be easily determined by the typical shooter with little more than a caliper and a reloading scale.

Before his death, Don Miller contributed some valuable ideas to the discussion of developing a formula for OTM bullets. We discussed possible ways to parameterize the open cavity in the bullet tip. One possibility was determining its volume by injecting water into the open cavity and measuring the increase in mass. Another possibility was inserting a needle or the lead of a mechanical pencil and subsequently measuring the depth of the cavity as the length of the lead or needle that was inserted. This method was attractive because it would be more easily executed by shooters and because it provided a way to think about the problem with analagous reasoning to the formula for plastic tipped bullets.

We reasoned that the Sg for OTM bullets would be less than the Sg predicted by the plastic tipped formula (using the total length minus the open tipped depth for the effective metal length) because the metal jacket at the tip of the bullet would contribute to the effective position of the center of mass and increase the tumbling moment of inertia more than the lighter plastic tip. We also reasoned that the Sg for OTM bullets would be greater than the Sg predicted by the original twist rule for solid bullets of near uniform density.

The authors of the present study lack Don Miller's passion for theoretical rigor and have taken a more empirical approach based on the above thinking. It would be reasonable to suppose that a good approximation of the stability of OTM bullets would be given by:

$Sg(OTM) = a * Sg(PT) + (1 – a) * Sg(CD),$    (Eqn. 1)

where Sg(OTM) is the stability of the OTM bullet, Sg(PT) is the predicted stability for a plastic tipped bullet (Courtney and Miller, 2012a), and Sg(CD) is the predicted stability of a constant density bullet (Miller, 2005; Miller, 2009). The parameter, a, is a weighting factor: a = 0.5 would be an evenly weighted average between the two earlier formulas; a = 0.9 would weight the average 90% of the way closer to the plastic tipped formula. Trial and error with available data (including some of the data below) and the expectation of reduced ballistic coefficients for Sg < 1.2 and tumbling near Sg = 1.0 suggests a = 0.7 so that a gyroscopic stability formula for OTM bullets is

$Sg(OTM) = 0.7 Sg(PT) + 0.3 Sg(CD)$    (Eqn. 2).

In the Sg(PT) formula, the length of the full density portion of the bullet is used for the "metal length." This can be computed as the total length minus the depth of the open tip portion as determined by inserting a thin drill bit, pencil lead, or other similar object.

**Method**

The experiment to measure ballistic coefficients as a function of Sg uses two CED Millenium chronographs with LED sky screens. The experimental care employed to ensure that the optical chronographs meet their specified accuracy level (0.3%) includes ensuring the skyscreen planes are parallel, adding mechanical rigidity to ensure minimal motion during the experiment (see Figure 1 of Courtney and Courtney, 2013), and keeping the projectile paths centered over the optical sensors in a window 50mm square.

The chronographs are further calibrated by placing them in the shot line, with minimal separation, and shooting though them. Each reading of the second chronograph is adjusted upward appropriately for the small loss of velocity (< 5 fps) over the two foot dis-

# Gyroscopic Stability of Open Tipped Match Style Rifle Bullets

tance from the closest chronograph. Then the average velocity of ten shots can be compared to determine systematic variations in the readings between the chronographs. In this manner, the variations between chronographs can be reduced to 0.1%.

After calibration, the chronographs are placed 10 feet and 160 feet from the muzzle. Chronograph separations are measured with a tape measure within a few inches. Near and far velocities are determined over the separation interval. Previous work (Courtney and Courtney, 2013) and the Sierra Reloading Manual (5$^{th}$ edition) suggest a 50 yard interval is most appropriate for measuring BCs sensitive to gyroscopic stability and coning motions. Environmental conditions were measured with a Kestrel 4500 pocket weather meter. BCs were computed with the JBM ballistics calculator. (http://www.jbmballistics.com/cgi-bin/jbm-bcv-5.1.cgi ) Gyroscopic stability was computed with Eqn. 2.

Some have expressed doubt regarding the accuracy and repeatability of drag measurements using inexpensive optical chronographs. However, the accuracy of this experimental method for determining balistic coefficients has been validated at 1% using an independent acoustic method to determine time of flight. The repeatability has been validated by measuring the ballistic coefficient of the same laboratory standard bullet on multiple occasions at a level of 1%. Further, other authors (Litz, 2009) have also found the same model of chronograph used here to be adequate for BC measurements accurate to 1%.

The repeatability of this experimental method was additionally validated by measuring the drag coefficients of a 40 grain varmint bullet over the whole supersonic range in both Louisiana and Colorado and determining that drag coefficients were independent of both air density and moving and recalibrating the chronographs and different locations and elevations (Courtney et al., 2014). Repeatability is further supported by obtaining identical drag coefficients of 8 different subsonic projectiles in different experimental trials in Colorado and Louisiana.

As described previously (Courtney and Miller 2012b), one can shoot bullets out of the same rifle barrel over a range of stabilities by varying the velocity with different powder charges. In .223 Remington it is convenient to use two near full power loads with Varget or H4895 followed by a sequence of reduced loads beginning with 14 grains of Blue Dot and lowering the powder charge in 1 grain steps until a velocity is reached where the Sg is close to 1.0 or the bullet approaches Mach 1. After hitting the far chronograph with a tumbling bullet (Courtney and Miller, 2012a) in an earlier experiment, greater care was taken not to go far below Sg = 1.0. In .222 Remington, it is convenient to use two near full power loads with Varget or H4895 followed by a sequence of reduced loads beginning with 12 grains of Blue Dot and lowering the powder charge in 1 grain steps until a velocity is reached where the Sg is close to 1.0 or the bullet approaches Mach 1.

The approach of varying Sg by varying velocity in the same barrel is preferred to shooting from barrels with different twist rates, because other work has shown that effects from different barrels other than bullet stability can cause significant differences in aerodynamic drag (Litz, 2009). Varying the velocity also allows a number of Sg values to be achieved over a range without needed a separate rifle barrel for each Sg. The possible confounding effect of BC variation with muzzle velocity is relatively small compared with the decrease of BC with Sg.

In each case, the measured BC (using the G1 drag model) is graphed vs. the predicted Sg(OTM) from Eqn. 2. The uncertainty in BC (error bars) is estimated as the standard error of the mean.

**Results**

Figure 1 shows a graph of G1 ballistic coefficient vs. Sg(OTM) for the 62 grain Berger Flat Base (BFB) bullet fired from a Savage Model 25 in .222 Remington as the muzzle velocity is lowered from 2900 ft/s down to 1450 ft/s. The atmospheric conditions were a pressure of 24.74 in Hg, a temperature of 32.8 F, and relative humidity of 82%. Ballistic coefficients were calculated by entering the near and far velocities, along with the environmental conditions (ambient pressure, humidity, and temperature) into the JBM ballistic calculator. Note the relatively constant BC for Sg above 1.2 and the decreasing BC as Sg is lowered from 1.2 toward 1.0. The blue line in the figure was determined from prior BC measurements for this bullet using a 1 in 12" twist barrel in a slightly less dense atmosphere. All of the BC measurements in this Sg range were between 0.244 and 0.246. The total length of this bullet was

# Gyroscopic Stability of Open Tipped Match Style Rifle Bullets

0.810" and the open tip depth was 0.14", giving a full density length of 0.670".

This decreasing BC as the Sg is lowered toward 1.0 is expected and is consistent with the descriptions in the 4[th] and 5[th] editions of the Sierra Reloading Manuals. However, also note the increase in BC at Sg = 1.151. This reminds one of similar increases in BC near Sg = 1.23 in earlier work (Courtney and Courtney, 2013), but given the large error bars, it is difficult to assess its significance. Perhaps the Sg value where the "sweet spot" occurs is lower in a bullet with a relatively large volume in the open tip. However, it is possible that the rise is not significant.

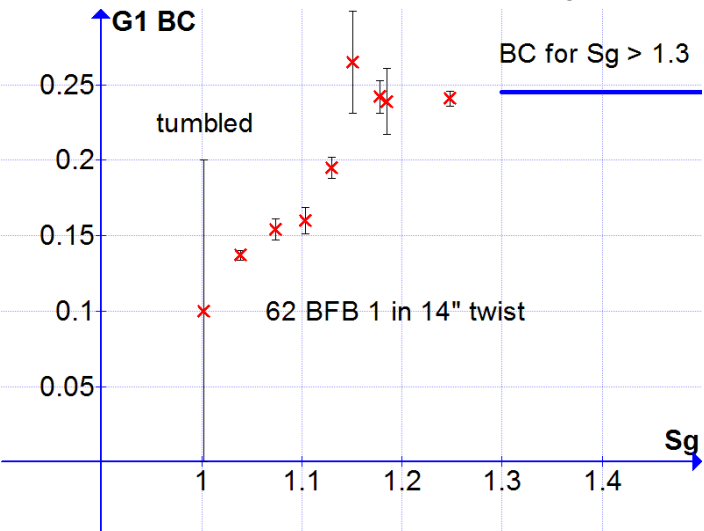

*Figure 1: Measured G1 ballistic coefficient vs. gyroscopic stability (OTM) of the 62 grain Berger Flat Base (BFB) open tipped match style bullet. Note the relatively constant BC above Sg = 1.2, the slight increase in BC near Sg = 1.15, and the decreasing BC as Sg is lowered from 1.2 to 1.0.*

The bullet was observed to tumble at Sg(OTM) = 1.022, supporting the suggestion that these empirical stability formulas tend to be accurate within 5% or so for bullets of the corresponding construction. The first shot at this stability failed to register on the far chronograph, which is not uncommon for tumbling bullets. The second shot actually hit one of the skyscreens on the far chronograph, always a risk in drag experiments employing downrange equipment. Tumbling by confirmed by observing keyholes in the target on subsequent shots.

It is notable that the original Miller twist rule (Miller, 2005; Miller, 2009) predicts Sg(CD) < 1.0 for the whole range of experimental data shown in Figure 1 for the 1 in 14" twist .222 Remington. Consequently, the empty space in the tip of the bullet significantly enhances (increases) stability and allows the bullet to be stable under conditions where a constant density bullet of the same dimensions would not be stable.

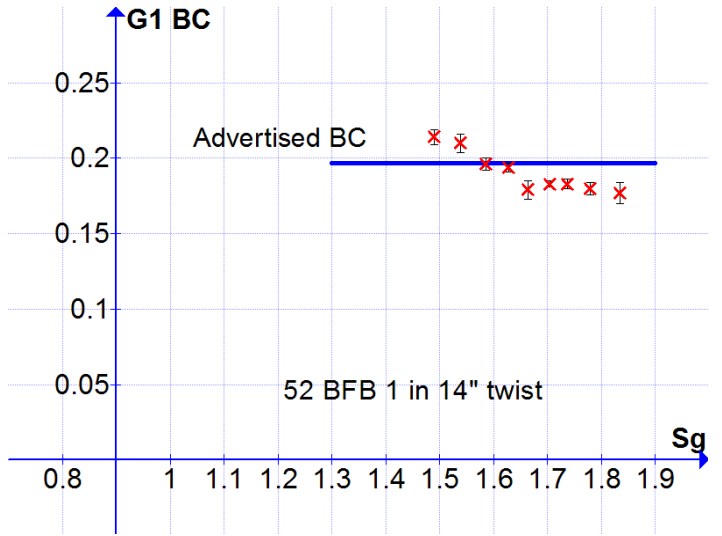

*Figure 2: Measured G1 ballistic coefficient vs. gyroscopic stability (OTM) of the 52 grain Berger Flat Base (BFB) open tipped match style bullet.*

Figure 2 shows a graph of G1 ballistic coefficient vs. Sg(OTM) for the 52 grain Berger Flat Base (BFB) bullet fired from a Savage Model 25 in .222 Remington as the muzzle velocity is lowered from 3120 ft/s down to 1430 ft/s. The atmospheric conditions were a pressure of 24.74 in Hg, a temperature of 34.0 F, and relative humidity of 81%. The original Miller stability formula predicts Sg(CD) ranging from 1.173 to 0.953 as the velocity is varied. This would suggest BCs much lower than predicted by the manufacturer and shown in Figure 2 for Sg > 1.0 and it suggests tumbling would be expected for the two least stable loads. The new formula predicts Sg(OTM) ranging from 1.835 to 1.490, which is more consistent with the absence of tumbling and the lack of decreasing BC as stability is lowered. Computation of Sg(OTM) used a total length of 0.718" and a full density length of 0.513" owing to an open tip depth of 0.205".

In experiments with different twist rate barrels, the 4[th] and 5[th] editions of the Sierra Reloading Manual report relatively constant BC for a number of different barrel

# Gyroscopic Stability of Open Tipped Match Style Rifle Bullets

twists until the stability begins to be marginal, in which cases, lower BCs are reported. Unfortunately, those reports do not contain information regarding the atmospheric conditions or actual estimates of Sg. The data in the Sierra manuals suggests that bullets do not suffer drag increases (BC reductions) by being overstabilized. That is there is no drag increase due to bullets remaining slighly canted in flight (not pointing exactly forward) as a result of higher stability.

However, the Sierra manuals do report slight drag increases for the hotter loads in some cartridges. The manuals suggest this may be due to higher gas pressure giving the bullet a larger tip off angle via forces on the bullet base as the bullet leaves the barrel. It is not clear how they eliminated increased stability as a possible confounding factor in these cases, since the experiments varying the twist rate were performed on different bullets and different cartridges.

The increase in drag as Sg(OTM) is increased from 1.490 to 1.835 is intriguing. The two highest Sg loads were near full power loads with 24 grains and 22 grains of H4895. QuickLoad predicts a muzzle pressure of 6631 psi for 24 grains of H4895, which is modest due to the modest case volume of the .222 Remington. The muzzle pressure for 22 grains of H4895 is predicted to be 5978 psi. If a high muzzle pressure leading to tip off were responsible for the increased drag at the higher stability, one would expect it to disappear for 10 to 12 grains of Blue Dot (Sg(OTM) of 1.663 to 1.763) corresponding to muzzle pressures from 3375 psi to 3841 psi. Yet all five of the highest Sg(OTM) loads have BCs from 0.177 to 0.183 with relatively small uncertainties. The BCs below the advertised BC for this bullet and below the measured BC for the lower range of Sg (1.490 to 1.628) seem more likely to due to the stability than to the muzzle pressure resulting in tip off.

Of course, repeating the BC measurements shown in Figure 2 for a higher (closer to sea level) air density would cover a lower range of stability for a comparable range of muzzle pressures and may assist both in further validation of the Sg(OTM) formula as well as determining whether the observed variation in BC is more likely due to stability or muzzle pressure.

Figure 3 shows a graph of G1 ballistic coefficient vs. Sg(OTM) for the 69 grain Nosler Custom Competition (NCC) bullet fired from a Remington 700 in .223 Remington (1 turn in 12" twist) as the muzzle velocity is lowered from 2870 ft/s down to 1800 ft/s. The atmospheric conditions were a pressure of 24.74 in Hg, a temperature of 34.0 F, and relative humidity of 81%. The original Miller stability formula predicts Sg(CD) ranging from 1.117 to 0.956 as the velocity is lowered. This suggests tumbling would be expected for the for the two least stable loads. The new formula predicts Sg(OTM) ranging from 1.278 to 1.094, which is more consistent with the absence of tumbling and the trend of decreasing BC as stability is lowered.

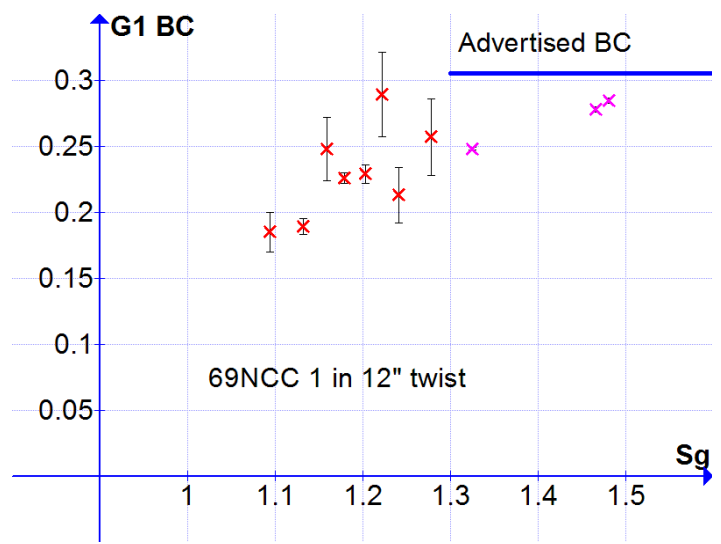

*Figure 3: Ballistic coefficient vs. gyroscopic stability, Sg(OTM), of the 69 grain Nosler Custom Competition (NCC). The red data points are from the present study. The pink data points are from earlier BC measurements over 100 yards in a slightly thinner atmosphere.*

The Sierra Reloading Manual (4[th] and 5[th] editions) note greater BC variability for longer bullets, and we certainly observed this for the 69 grain custom competition with a total length of 0.890" and a full density length of 0.805". The relatively large uncertainties (due to shot to shot BC variations) for Sg(OTM) < 1.3 make detailed observations difficult beyond noting the trend for BC to decrease as Sg decreases. The trend of BC decreasing with Sg(OTM) as well as the lack of bullet tumbling over the range support the reliability of the new formula for predicting Sg(OTM).

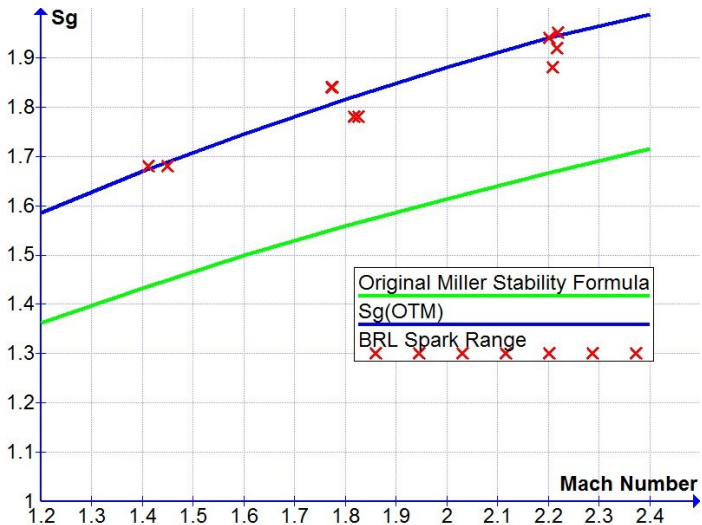

*Figure 4: Comparisons of supersonic BRL spark range gyroscopic stability data (McCoy, 1988) compared with the new stability formula for open tipped match bullets, Sg(OTM) as well as the original Miller stability formula, Sg(CD). Bullet is the 168 grain 0.308" Sierra International.*

Data acquired at the BRL spark range provides opportunity to compare experimentally determined Sg values with stability formula predictions for the 168 grain Sierra International bullet in .308. Figure 4 shows the supersonic data from BRL (McCoy, 1988) compared with predictions of the original Miller stability formula (Miller, 2005) and the Sg(OTM) formula (Eqn. 2). It is clear that the original formula underestimates the bullet stability (as expected), and that the Sg(OTM) formula provides good agreement. In the supersonic range (above M1.2), predictions from the Sg(OTM) formula range from 3.4% high to 1.8% low, with a root mean square (rms) error of 1.8%. In contrast, the original Miller stability formula underestimates the experimentally determined bullet stability by an average of 13.6% in the supersonic range.

The original experimental data presented above is limited to the supersonic regime (above M1.2), but spark range data for the 168 grain Sierra International was included in McCoy (1988). Figure 5 shows the comparison of the BRL data with the Sg(OTM) and original Miller stability formula for constant density bullets. Figure 5 shows that the gyroscopic stabilities are higher than the predicted values in the transonic and subsonic regions. The Sg(OTM) formula makes more accurate predictions than the original Miller stability formula, but neither approach is particularly accurate. However, both formulas tend to underestimate stability in the subsonic and transonic region.

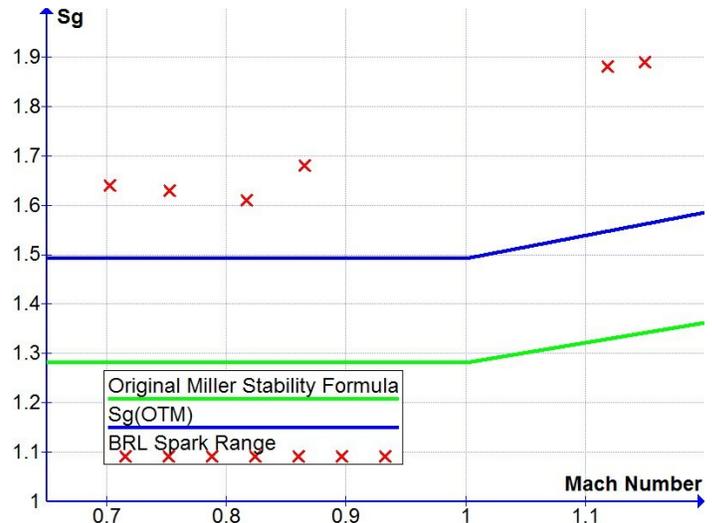

*Figure 5: Comparisons of subsonic and transonic BRL spark range gyroscopic stability data (McCoy, 1988) compared with the new stability formula for open tipped match bullets, Sg(OTM) as well as the original Miller stability formula, Sg(CD). Bullet is the 168 grain Sierra International in .308.*

**Discussion**

The experimental results presented above show that the original Miller twist rule (referred to as Sg(CD) here, referring to the assumption of constant density) tends to underestimate the gyroscopic stability of open tipped match style bullets and predicts instability for bullets and cases where, in fact, stable and do not tumble. The new formula for the gyroscopic stability of open tipped match style rifle bullets, Sg(OTM), is shown to be consistent (within 5%) with the experimental data reported here for supersonic muzzle velocities, accurately predicting decreases in BC for Sg(OTM) < 1.2 and correctly predicting the potential to tumble within 5% of Sg(OTM) = 1.00. The experimental results here raise new questions regarding possible drag rise for Sg > 1.6.

There are now applicable formulas for predicting gyroscopic stability for the three dominant rifle bullet styles: 1) near constant density jacketed lead or solid copper bullets without pronounced open tips, hollow points, or plastic tips 2) plastic tipped bullets and 3) open tipped match style bullets. The typical accuracy of these empirical formulas is expected to be within

# Gyroscopic Stability of Open Tipped Match Style Rifle Bullets

5% for most cases. There are not applicable Sg formulas for hollow point rifle bullets or for some of the composite bullets where the composite inside of the jacket is much less dense than the jacket material.

The applicability of the new formula for OTM bullets should probably be limited to cases where the depth of the open tip is 30% or less than the total length. And while 5% accuracy is expected for supersonic muzzle velocities, stability should always be verified at a shooting range under similar conditions to the actual application before relying too heavily on performance in a given application.

Other than the BRL spark range at Aberdeen Proving Ground (Maryland), the most definitive test of formulas predicting gyroscopic stability is whether or not these formulas accurately predict tumbling. Studying changes in BC as stability is changed provides additional insight, but should not be considered definitive.

Care should be exercised in distinguishing gyroscopic stability from dynamic stability. Gyroscopic stability governs how a bullet flies and whether it remains point forward in early flight (muzzle to a few hundred yards or so). Bullets such as the 168 grain SMK and the M855 are known to encounter dynamic instability problems at long range, even though they have adequate gyroscopic stability. A discussion of dynamic stability is beyond the scope of the current paper, but one cannot reasonably infer anything about long range dynamic stability from gyroscopic stability.

## Acknowledgements
This work was funded by BTG Research (www.btgresearch.org ) and the United States Air Force Academy. The authors are grateful to Don Miller for his encouragement on the project and to Colorado Rifle Club where the experiments were performed. The authors appreciate valuable feedback from several reviewers which has been incorporated into revisions of the manuscript.